\documentclass{ws-p9-75x6-50}

\begin{document}

\title{Axisymmetric electrovacuum spacetimes
with an additional Killing vector and radiation}

\author{A. Pravdov\' a}

\address{Mathematical Institute, 
Academy of Sciences, 
\v Zitn\' a 25,
115 67 Prague 1, Czech Republic \\E-mail: { 
 pravdova@math.cas.cz}}

\author{J. Bi\v c\' ak}

\address{Institute of Theoretical Physics, Faculty of Mathematics
and Physics, 
Charles University, V Hole\v sovi\v ck\' ach 2, 
180 00 Prague 8, Czech Republic\\E-mail: {
 bicak@mbox.troja.mff.cuni.cz}}

\def \a {\alpha}
\def \b {\beta}
\def \g {\gamma}
\def \G {\Gamma}
\def \d {\delta}
\def \eps {\varepsilon}
\def \ep {\epsilon}
\def \e {\eta}
\def \f {\phi}
\def \ffi {\varphi}
\def \j {\iota}
\def \th {\theta}
\def \vth {\vartheta}
\def \k {\kappa}
\def \l {\lambda}
\def \m {\mu}
\def \n {\nu}
\def \x {\xi}
\def \p {\pi}
\def \r {\rho}
\def \s {\sigma}
\def \t {\tau}
\def \ps {\psi}
\def \o {\omega}
\def \z {\zeta}
\def \L {\pounds}
\def \vu {\tilde u}
\def \der {\partial }
\def \nn {\nonumber}
\def \rov {\equiv}
\def \A {{\cal A}}
\def \BB {{\cal B}}
\def \C {{\cal C}}
\def \D {{\cal D}}
\def \E {{\cal E}}
\def \F {{\cal F}}
\def \GG {{\cal G}}
\def \K {{\cal K}}
\def \N {{\cal N}}
\def \L {{\cal L}}
\def \cN {\bar\N}
\def \vN {\tilde\N}
\def \M {{\cal M}}
\def \vM {\tilde \M}
\def \I {{\cal I}}
\def \J {{\cal J}}
\def \R {{\cal R}}
\def \CC {{\rm C}}
\def \U  {{\cal U}}
\def \T  {{\cal T}}
\def \bfi {\bar \varphi}
\def \br {\bar \rho}
\def \bz {\bar z}
\def \bt {\bar t}
\def \pul {{{\scriptstyle{\frac{1}{2}}}}}
\def \sn {\sin \th}
\def \cs {\cos \th}
\def \tg {\tan \th}
\def \ctg {\cot \th}
\def \csec {\csc \th}
\def \dcsec {\csc^2 \th}
\def \msn {\sin^{-1} \th}
\def \ct {\bar t}
\def \dsn {\sin^2 \th}
\def \mdsn {\sin^{-2} \th}
\def \tsn {\sin^3 \th}
\def \csn {\sin^4 \th}
\def \dcs {\cos^2 \th}
\def \tcs {\cos^3 \th}
\def \csin {\sqrt{1-(wu)^2}}
\def \dctg {\cot^2 \th}
\def \chd  {\cosh 2\d}
\def  \shd  {\sinh 2\d}
\def \mm {\mbox{\quad }}
\def \mv {\mbox{\qquad }}
\def \msip {\rightarrow}
\def \vsip {\longrightarrow}
\def \lkz  {\bigl(}
\def \pkz  {\bigr)}
\def \lvkz {\Bigl(}
\def \pvkz {\Bigr)}
\def \lvvkz {\biggl(}
\def \pvvkz {\biggr)}
\def \lhz  {\bigl[}
\def \phz  {\bigr]}
\def \lvhz {\Bigl[}
\def \pvhz {\Bigr]}
\def \lvvhz {\biggl[}
\def \pvvhz {\biggr]}
\def \lsz   {\bigl\{ }
\def \psz   {\bigr\} }
\def \pvsz {\Bigl\} }
\def \lvsz {\Bigr\{ }
\def \lvvsz {\Biggl\{}
\def \pvvsz {\Biggr\}}
\def \BE {\begin{equation}}
\def \EE {\end{equation}}
\def \BDM {\begin{displaymath}}
\def \EDM {\end{displaymath}}
\def \BEAH {\begin{eqnarray*}}
\def \EEAH {\end{eqnarray*}}
\def \BEA {\begin{eqnarray}}
\def \EEA {\end{eqnarray}}
\def \BM {\begin{math}}
\def \EM {\end{math}}
\def \BDM {\begin{displaymath}}
\def \EDM {\end{displaymath}}
\def \mm {\mbox{\quad }}
\def \mv {\mbox{\qquad }}
\def \msip {\rightarrow}
\def \vsip {\longrightarrow}
\def \Fab {F_{\a \b }}
\def \Fmn {F_{\m \n }}
\def \Fgb {F_{\g \b }}
\def \Fag {F_{\a \g }}
\def \Fnn {F_{00}}
\def \Fnj {F_{01}}
\def \Fnd {F_{02}}
\def \Fnt {F_{03}}
\def \Fjn {F_{10}}
\def \Fjj {F_{11}}
\def \Fjd {F_{12}}
\def \Fjt {F_{13}}
\def \Fdn {F_{20}}
\def \Fdj {F_{21}}
\def \Fdd {F_{22}}
\def \Fdt {F_{23}}
\def \Ftn {F_{30}}
\def \Ftj {F_{31}}
\def \Ftd {F_{32}}
\def \Ftt {F_{33}}
\newcommand{\zl}[2]{{{\scriptstyle{\frac{#1}{#2}}}}}
\def \pul {{{\scriptstyle{\frac{1}{2}}}}}
\def \tripul {{{\scriptstyle{\frac{3}{2}}}}}
\def \ctvrt {{{\scriptstyle{\frac{1}{4}}}}}
\def \osmina {{{\scriptstyle{\frac{1}{8}}}}}
\def \sestina {{{\scriptstyle{\frac{1}{6}}}}}
\def \VUW {\left(\V\edb-r^2\edg U^2\chd
           -r^2\emdg W^2\chd-2r^2 UW\shd\right)}
\def \V  {\frac{V}{r}}
\def \edb  {{\rm e}^{2\b }}
\def \edg  {{\rm e}^{2\g }}
\def \emdg {{\rm e}^{-2\g }}
\def \cc {(c,_\th+2c\ctg)}
\def \dd {(d,_\th+2d\ctg)}
\def \gab {g_{\a \b }}
\def \gmn {g_{\m \n }}

\maketitle

In the~present note we briefly summarize our recent work\cite{ajajibi,ajajibi2}
on possible additional symmetries of axially symmetric 
electrovacuum spacetimes 
which admit radiation.
The~main result  states that only boost and rotation (axially) symmetric
electrovacuum spacetimes can be radiative
and asymptotically flat at null infinity $\J$ which admits global sections.
If an additional symmetry is a translational spacelike or null Killing field
the~spacetime represents cylindrical or plane-type waves, local $\J$ may still exist
but some of its generators are missing.
Boost-rotation symmetric spacetimes are the~only 
known exact explicit radiative solutions of Einstein's equations
describing moving objects -- singularities or black holes uniformly
accelerated along the~axis of symmetry. They are radiative and admit
a smooth $\J$ although at least four points of $\J$ are missing.
They represent the~only known 
examples in which {\it arbitrarily strong} initial data with the~given
symmetry can be chosen on a hyperboloidal hypersurface which evolve into 
a complete, smooth null infinity and regular timelike infinity.
For the~latest reviews, containing a number of relevant references,
see  [3, 4]. 

Most recently the~analysis of asymptotic symmetries was extended
to spacetimes with null dust\cite{ajauta} and to spacetimes with
polyhomogeneous $\J$\cite{Kroon}.

In Refs. [1, 2]   
we assume that in addition to bound systems
and radiation, the~axisymmetric spacetimes may contain an infinite
(cosmic) string along the~axis of symmetry. Here we assume spacetimes
without strings. Our main results can then be summarized in the~following
three theorems, the full versions of which and detailed proofs are given in 
Refs.~[1] and [2].   

{\bf Theorem 1}\ {\it Suppose that an axially symmetric
electrovacuum spacetime admits a ``piece'' of $\J^+$ 
in the~sense that it admits  the~Bondi-Sachs expansions 
of the~metric and electromagnetic field in coordinates
 \mbox{ $\{ u$,~$r$,~$\th$,~$\f$\}}. 
Assume the~spacetime admits an additional Killing
vector $\e$ 
 forming with the~axial Killing vector a two-dimensional Lie
algebra (Killing vectors need not be hypersurface orthogonal)
and the~electromagnetic field  has the~same symmetry.
Then the~additional Killing vector has asymptotically
the~form 
\BE
\e^\a=[-ku\cs+\a(\th),\
          kr\cs+{\cal O}(r^{0}),\ -k\sn+{\cal O}(r^{-1}),\ 
{\cal O}(r^{-1})]\ ,  \label{Bbotr}
\EE
where $k$ is a constant and  $\a$ 
an arbitrary function of $\th$.
For $k=0$ it generates asymptotically translations.
For $k\not= 0$ it is the~boost Killing field
(one can put, without loss of generality, $\a=0$ and $k=1$)
which generates the~Lorentz transformations
along the~axis of axial symmetry.}\\


{\bf 1) Translational Killing vectors:} 

{\bf Theorem 2} {\it If an axisymmetric electrovacuum spacetime 
admits a local $\J^+$ and 
an asymptotically spacelike translational Killing vector,  
and if  the~spacetime is not flat in the~neighbourhood of $\J^+$ 
(i.e. 
the~Bondi mass aspect $M$ or  other metric functions
are non-vanishing)
then $\J^+$ has singular generators at 
$\th=\th_0\not= 0,\ \p$ 
and the~spacetime contains cylindrical waves. 
If the~translational Killing vector is null, 
then $\J^+$ is singular at $\th=0$ or $\p$ --
a wave propagating along the~symmetry axis is present. 
}

{\bf Theorem 3}
{\it If an axisymmetric electrovacuum
spacetime 
with a non-vanishing Bondi mass $m$
admits an asymptotically translational Killing vector
and a complete cross section of $\J^+$, then the~translational
Killing vector is timelike and spacetime is thus stationary.}\\


{\bf 2) The~boost Killing vector:} 

We now find non-vanishing news functions, $c,_u$, $d,_u$ 
(describing gravitational radiation) and  $X$,  $Y$ 
(describing electromagnetic radiation)
to have the~forms:
\BE
c,_u(u,\th)\! =\!\frac{K(w)}{u^2},\ 
d,_u(u,\th)\! =\!\frac{L(w)}{u^2}, \ 
X    (u,\th)\! =\!\frac{\E(w)}{u^2},\ 
Y    (u,\th)\! =\!\frac{\tilde{\BB}(w)}{u^2},
\EE
where $K$ etc are functions of $w=\sn /u$. They determine the~mass aspect $M$, 
other metric and field functions and the~Bondi mass:
\BE
M(u,\th)\! =\!\frac{(w^2K),_w}{2\sn}+\frac{\l(w)}{w^3u^3} 
\EE
where
\BE
\l,_w\! =\!w^2(K^2+L^2+\E^2+\tilde{\BB}^2)
-\frac{(w^3K,_{w}),_w}{2w}\ .\label{boostlambda}
\EE
The~Weyl tensor
has, in general, radiative components ($\sim 1/r$). 
The~boost-rotation symmetric  spacetimes are clearly 
radiative\cite{JibiEhlers,AVreview}.

\section*{Acknowledgments}
\mbox{A. P.} acknowledges  grants \mbox{{GA\v CR} 201/98/1452},
\mbox{{GA\v CR} 202/00/P031},  \mbox{J. B.} grants
\mbox{{GA\v CR} 202/99/0261}, \mbox{GAUK 141/2000}.

\end{document}